\documentstyle[twocolumn,prc,aps,epsfig,psfig]{revtex}

     \newcommand{\pathnow}{}
\begin{document}\hbadness=10000\topmargin -1cm
  \hyphenation{strang-en-ess}
\twocolumn[\hsize\textwidth\columnwidth\hsize\csname %
@twocolumnfalse\endcsname
\title{Search for QGP and Thermal Freeze-out of Strange Hadrons 
}
\author{Giorgio Torrieri and Johann Rafelski}
\address{
Department of Physics, University of Arizona, Tucson, AZ 85721
}
\date{November, 2000}
\maketitle
\begin{abstract}
After reviewing the observables of QGP we perform an analysis
of $m_\bot$ spectra of strange hadrons measured as function
of centrality in 156$A$\,GeV Pb--Pb interactions. We show that 
there is a good agreement between the chemical and thermal
freeze-out conditions, providing additional evidence for 
the formation and  sudden disintegration of a super cooled 
QGP fireball.\\

PACS: 12.38.Mh, 12.40.Ee, 25.75.-q
\end{abstract}
\pacs{PACS: 12.38.Mh, 12.40.Ee, 25.75.-q}
]
\begin{narrowtext}
\section{Quark Matter in Laboratory}
The development of the quark model  has been from the 
first moments  accompanied by consideration of the 
transition from a few body hadronic
bound state to a many body quark matter star 
formation \cite{Ito70}. This was followed by
the development of the  quantum many body theory of 
quark matter \cite{Car73,ILL74}, which lead on to the
formal recognition within the framework of QCD that
perturbative quark matter state must exist \cite{Col75},
given the asymptotically free nature of theory of strong 
interactions, quantum-chromodynamics (QCD). 
Arguments arising from study of a dense
 hadron gas within the scheme of
Hagedorn's statistical  bootstrap and the resulting 
boiling of hadronic matter lead from a different direction
to the consideration of the transition to
a hadron substructure phase \cite{Hag84}. 
In short, the quark-gluon plasma (QGP) as we today 
call hot quark matter
has been for 30 years an expected  new state of matter.

The nuclear physics establishment considered at that time,
if not still today, other ideas about new phases of matter which could
be  formed in relativistic nuclear collisions, to be of 
greater interest.  It is interesting to recall here that in
the first of the series of formative workshops
in the field of relativistic heavy ion collisions,
the ``Bear Mountain'' meeting convoked in November 1974 there is 
not a single  mention of quarks, let alone of quark matter.
 At  the time ``Lee-Wick'' ultra dense nuclei, pion condensates,
multi-hyperon nuclear states were considered as 
the discovery potential of these new and coming tools of 
nuclear physics research. 

As the ideas about QGP formation in high energy nuclear 
collisions matured \cite{Chi78}, a challenge  emerged how
the locally deconfined state which exists a mere 10$^{-22}$s
can be distinguished from the gas of confined hadrons. This is 
also a matter of principle, since arguments were advanced 
that this may be impossible since both quark and hadron pictures 
of the reaction are equivalent. Therefore a quark-gluon
 based description is merely a change of Hilbert space 
expansion basis.

Clearly these difficult questions can be settled by an 
experiment, if a probe of QGP operational on the collision
time scale, can be devised. There were 
three major groups of  observables proposed, and we
address these in the chronological order of their
appearance in literature.

\underline{Dileptons, direct Photons}: The study of multi particle
production phenomena has stimulated the exploration of 
dileptons and photons as the probe of dense hadron matter fireballs,
which ideas were easily adapted to the QGP phase situation. 
After the seminal work of Feinberg \cite{Fei76} and Shuryak \cite{Shu78}
a comprehensive discussion of this observable was offered  \cite{Dom81}. 

However, since electromagnetic currents are the source 
of photons and dileptons,
both confined and deconfined dense elementary hadron matter can produce 
these electromagnetic   probes. The principal novel component of QGP, 
gluons, are not a required ingredient. 

On practical side, the  backgrounds are very significant. The
photon production is dominated by factor 10-20 larger
$\pi^0$ decay, and dileptons arise in decays of vector mesons
which are also abundantly produced in multi particle 
production processes, irrespective if the formation of QGP
has occurred or not. Thus  electromagnetic 
signature of QGP has to be extracted  
comparing in a detailed and quantitative study experiment with theory.

Such a comparison is extremely difficult unless we we have good data 
and already know at what condition QGP has been formed and how it 
evolved. But present day experiments suffer both from 
systematic acceptance issues, and low statistics. In our view, the
electromagnetic probes will come of age as a second generation diagnostic 
tools in the refinement of the study of the QGP phase properties. 

\underline{Strangeness enhancement}:
One  aspect of this probe of QGP will be addressed in
this paper. The ideas about enhancement are simple: 
when color bonds are broken the chemically (abundance) equilibrated 
deconfined state has an unusually high abundance of strange
quarks \cite{abundance}. Subsequent study of the dynamical process 
of chemical equilibration  has shown that only the  
gluon component in the QGP is 
able to produce strangeness rapidly \cite{RM82}, allowing formation
of  (nearly) chemically equilibrated dense phase of deconfined,
hot, strangeness-rich, quark matter in relativistic nuclear collisions. 
Therefore strangeness enhancement is related directly to 
presence of gluons in QGP.

The high density of strangeness formed in the reaction fireball 
favors formation of multi strange hadrons \cite{Raf82,RD83}, which 
are produced  rarely if only individual hadrons 
collide \cite{Koc85,KMR86}. In particular a large enhancement 
of multi strange antibaryons 
has been proposed as  characteristic and nearly background-free 
signature of QGP \cite{Raf82}. Such a systematic enhancement has
in fact been observed, rising with strangeness content \cite{WA97p}. 

Although conventional theoretical models were explored 
to interpret the strangeness signatures of new physics \cite{WA97gen}, 
we are not aware of a consistent interpretation of the
data other then in the context of QGP formation. Experimental results 
are abundant and allow a precise diagnosis
of the chemical freeze-out conditions \cite{Let00}, 
and an assessment about the initial
conditions \cite{Ham00}.

\underline{Charmonium}:
By the time the CERN experiment NA38 was  taking data on Charmonium 
production in nuclear collisions, the possibility that formation 
of QGP will influence  the final state yield of  charmonium  has been 
raised \cite{Mat86}. This important topic
has undergone a significant evolution over the  past 15 years. In fact
the originally predicted charmonium suppression at SPS 
energy can turn to a charmonium enhancement  at RHIC energies. 
By virtue of detailed balance, both dissociation and 
QGP based charmonium production channels must be considered.
A requirement  for dominating enhancement effect is a sufficient number 
of available open charm quark pairs in the QGP \cite{The00}.

Based on these  diverse experimental signatures,
it is believed that a new form of  matter,
presumably quark matter, 
has been formed in  relativistic nuclear collision 
experiments carried out at CERN-SPS \cite{CERN}.  Detailed 
analysis of  strangeness production results in addition 
implies that a dense fireball of matter formed in these
reactions  expands explosively, super cools, 
and in the end encounters a mechanical instability which facilitates 
sudden break up into hadrons  \cite{Raf00}. 

For the study presented in this paper 
the key point is that  when sudden QGP breakup occurs, 
the  spectra of hadrons are not formed at a range of 
stages in fireball  evolution, but arise rather
suddenly. Most importantly, particles of very different 
properties are produced by the same mechanism and thus are 
expected to have similar
$m_\bot$-spectra as is indeed observed \cite{Ant00}.  
The reported symmetry of the strange
baryon and antibaryon spectra is strongly suggesting that 
the same reaction mechanism produces $\Lambda$ and $\overline\Lambda$ 
and $\Xi$ and $\overline\Xi$. 
This is a surprising, but rather clean
experimental fact, which we will 
interpret quantitatively in this paper. 

When the momentum distributions of final state particles 
stop evolving during the fireball evolution, we speak of thermal
freeze-out. Because a spectrum of strange hadrons includes
directly produced, and heavy resonance decay products,
one can determine the freeze-out temperature and 
dynamical velocities of fireball evolution 
solely from the study of  precisely known 
shape of the particle spectra. We demonstrate
this in some detail in section 
\ref{chidata}. The physical mechanism is that 
the freeze-out temperature  determines the 
relative contribution  of each decaying resonance
while the shape of each  decay contribution differs 
from the thermal shape, see section \ref{Thermal}. 

We note that we 
make in our analysis the tacit assumption that 
practically all decay products of resonances 
are thermally not re-equilibrated, which is equivalent 
to the assumption of sudden freeze out. This is consistent 
with our finding that  the $m_\bot$ strange baryon 
and antibaryon  distributions of $\Lambda, 
\overline\Lambda, \Xi, \overline\Xi$ froze out 
near to the condition at which the chemical
 particle yields were  established. 

One of the key objectives of this  work is to present 
a comparison between thermal 
and chemical freeze-out analysis results for temperature,
(explosion) collective velocity and other chemical 
and dynamical parameters. It is important to realize that 
particle spectra and yields are sensitive to magnitude of 
collective matter flow, in which 
produced particles are born, for somewhat different 
reasons:
1) in thermal analysis the collective flow combines with
thermal freeze-out temperature to fix the shape of each particle 
spectrum, and temperature is also controlling the relative yield of 
contributing resonances -- see previous paragraph for 
a here relevant tacit assumption -- thus both $T$ and $v$ 
are fixed by  the shape of $m_\bot$ data; 
2) in chemical analysis the  particle yields required are obtained
integrating spectral yields, with experimental 
acceptance in $p_\bot,y$ implemented \cite{Let00}. Since many
particles have a too small  particle momentum to be
usually observed,
the acceptance-cut yields used in chemical analysis 
depend quite sensitively
on parameters which   deform the soft part of the 
spectra without changing
the number of produced particles, such as is the flow velocity.
For this reason precise particle spectra and yields
are allowing to draw conclusions about the proximity of 
thermal and chemical 
freeze-out conditions.

\section{Thermal freeze-out analysis}\label{Thermal}
In recent months experiment WA97 
determined the absolute normalization of
 the published 
$m_\bot$ distribution \cite{Ant00},  and we
took the opportunity to perform the spectral shape analysis
and will compare  our results to those obtained in chemical
yield analysis \cite{Let00} in order to check if the 
thermal and chemical freeze-out conditions are the same.
Our analysis continues and this report gives its current status.

We report here a simultaneous analysis of absolute yield and shape of 
WA97 results of six $m_\bot$-spectra of $\Lambda,\,\overline\Lambda, \,\Xi,\,
\overline\Xi,\, \Omega+ \overline\Omega, \, K_s=(K^0+\overline{K^0})/2$
in four centrality bins.
If thermal and chemical freeze-outs are identical, our
present results  must  be consistent with 
earlier chemical analysis of hadron yields. Since
the experimental data we here study 
is dominated by the shape of $m_\bot$-spectra
and not by relative particle yields, our analysis is de facto 
comparing thermal and chemical freeze-outs. 

We have found, as is generally believed and expected, 
 that all hadron $m_\bot$-spectra are strongly influenced by 
resonance decays. Thus we apply standard procedure to 
allow for this effect \cite{Sch95,Ani85}. The final
particle distribution is composed of directly produced particles
and decay products:
\begin{equation}\label{2body}
\frac{dN_X}{dm_\bot} =  \frac{dN_X}{dm_\bot}\vert_{\scriptsize\rm direct} +
\!\!\!\! \sum_{ \forall R \rightarrow X + 2+... } 
\frac{dN_X}{dm_\bot}\vert_{R \rightarrow X + 2 +\ldots}  
\end{equation}
Here:
$R(M,M_T,Y) \rightarrow X(m,m_T,y)+2(m_2)+\ldots$, where we 
indicate by the arguments that only for the decay  particle  $X$ 
we keep the information about the shape of the
momentum spectrum. We consider here only the 2-body 
decay as no other contributing decays are known for hyperons, and 
hard kaons. In detail, the decay contribution to yield of 
$X$ is:
\begin{eqnarray}
\label{reso}
\frac{dN_X}{d {m^2_\bot} d y }&=&
\frac{g_{r} b}{4 \pi p^{*}}
\int_{Y_-}^{Y_+}\!\! dY
\int_{M_{T_-}}^{M_{T_+}} dM_{T}^{2}\, J \,
\frac{d^2 N_{R}}{dM_{T}^{2} dY}  
\\[0.3cm]\nonumber
J&=&\frac{M}{\sqrt{P_{T}^2 p_{T}^2 -\{M E^{*} - M_{T} m_{T} \cosh\Delta Y\}^2}}
\end{eqnarray}
We have used $\Delta Y=Y-y$, and
$\sqrt{s}$ is the combined invariant mass of the 
decay products other than particle $X$
and $E^{*}=(M^2-m^2-m_2^2)/2M$, $p^{*}=\sqrt{E^{*2}-m^2}$ 
are the energy and momentum of the decay
particle X in the rest frame of its parent.
The limits on the integration are the maximum values accessible
to the decay product $X$:
\[\ Y_{\pm}=y \pm \sinh^{-1}\left(\frac{p^{*}}{m_{T}}\right) \]
\[\ M_{T_{\pm}}=M 
\frac{E^{*} m_{T} \cosh\Delta Y \pm p_{T} 
\sqrt{p^{*2}-m_{T}^{2} \sinh^{2} \Delta Y}}
{m_{T}^{2} \sinh^{2} \Delta Y+m^{2}}\]

The theoretical primary particle
spectra (both those directly produced and parents of 
decay products) are derived from the Boltzmann 
distribution by Lorenz-transforming from a flowing 
intrinsic fluid element to the CM-frame, and 
integrating over allowed 
angles between particle direction  and local flow.

We introduce two velocities: a local  flow velocity $v$ of fireball 
matter where from particles emerge,
and hadronization surface (breakup) velocity which we refer to 
as $v_f^{\,-1}\equiv dt_f/dx_f$.  Particle production is controlled by the 
effective volume element, which comprises 
\begin{equation}\label{v2v}
d S_{\mu} p^{\mu} = 
  d \omega \left(1- \frac{\vec v_{f}^{\,-1} \cdot \vec p}{E}\right)\,,
\qquad  d\omega \equiv \frac{d^3xd^3p}{(2\pi)^3}\,.
\end{equation}
The  Boltzmann distribution we adapt has thus the form
\begin{equation}
\frac{d^2 N}{dm_{T} dy} \propto
\left(1- \frac{\vec v_{f}^{\,-1} \cdot \vec p}{E}\right)
\gamma\,  m_{T} \cosh y \,
e^{-\gamma \frac E T \left(1-\frac{\vec v\cdot \vec p}{E}\right)}\,,
\end{equation}
where $\gamma=1/\sqrt{1-v^2}$\,.

The normalization for each hadron type $h=X,R$ is
\[\ N^h = V_{\scriptsize\rm QGP} \prod^{n}_{i\in h} \lambda_{i} \gamma_{i} \]
Here we use the chemical parameters 
$\lambda_{i}, \gamma_{i}$ $i=q,s$ which  are as defined in \cite{Let00}.
and commonly used to characterize relative and absolute abundances of 
light and strange quarks.

Since particle spectra we consider have a good relative
normalization, only one parameter is required for 
each centrality in order to describe the absolute normalization
of all six hadron spectra. This is for two reasons important:\\
a) 
we can check if the volume from which strange hadrons 
are emitted grows with centrality of the collision as
we expect;\\
b) 
we can determine which region in $m_\bot$ 
produces the excess of $\Omega$  noted in the
chemical fit \cite{Let00} is coming from.

However, since the normalization $V_{\scriptsize\rm QGP}$ common for
all particles at given centrality comprises additional  
experimental acceptance normalization, we have  not 
determined the value of the fireball emission volume at each centrality. 
Hence we will be presenting the volume parameter as
function of centrality  in arbitrary units.

The best thermal and chemical  parameters which minimize the total relative
error $\chi^2_{\rm T}$ at 
a given centrality:
$$\chi^2_{\rm T}=\sum_i\left(\frac{F^{\scriptsize\rm \,theory}_i-F_i}
                        {\Delta F_i}\right)^2\,,$$
for all experimental measurement points $F_i$  having
measurement error $\Delta F_i$
are determined by considering simultaneously for results of experiment WA97 \cite{Ant00} 
i.e. K$^{0}$, $\Lambda$, $\overline{\Lambda}$, $\Xi$, $\overline{\Xi}$,
$\Omega + \overline{\Omega}$. We have checked the validity of the 
statistical analysis by the usual method, i.e. omission of 
some data in the fit. 

Only in case of Kaons we find any impact of such a
procedure. Noting that the statistical error of kaon spectra 
is the smallest, we have established how a a systematic
error which could be for Kaons greater than statistical 
error would influence our result.  For this purpose 
we assign to   K$^0$ experimental results in most of our analysis  
an `error'  which  we arbitrarily  have chosen to be 5 times greater 
than the statistical error. In this way the weight of the kaon spectra 
in the analysis is greatly reduced. In the first result figure below 
(figure \ref{TdT}) we present both results, standard K$^0$ error and enlarged 
error. We see that while in individual result some change can occur,
overall the physical result of both analysis are consistent. Thus we 
can trust in the combined study of hyperon and kaon data. This conclusion
is reaffirmed in section \ref{chidata}, where we will see that the 
minimization of $\chi^2_{\rm T}$ involve more or less pronounced 
minima, depending on the error size of the kaon spectra,
see Fig.\,\ref{TchiT2}. In most calculations we present in this
paper we will be using, unless otherwise said, hyperon results combined
with the Kaon data with 5 times enlarged statistical error. We believe
that in this way we will err on the conservative side in our 
physical conclusions.

\section{Overview of the Results}
We show here a slate of results obtained within the 
approach outlined above. First we address the parameters
determining the shape of the $m_\bot$ distributions,
that is $T,v,v_f$.

As function of the centrality bin we show in Fig.\,\ref{TdT}
the freeze-out temperature $T$ 
of the $m_\bot$ spectra. The horizontal lines 
delineate range of result of the most recent  chemical 
 freeze-out analysis , see  Ref. \cite{Let00}.
\begin{figure}[tb]
\vspace*{-1.3cm}
\centerline{
\psfig{width=9.cm,clip=,angle=-90,figure=\pathnow 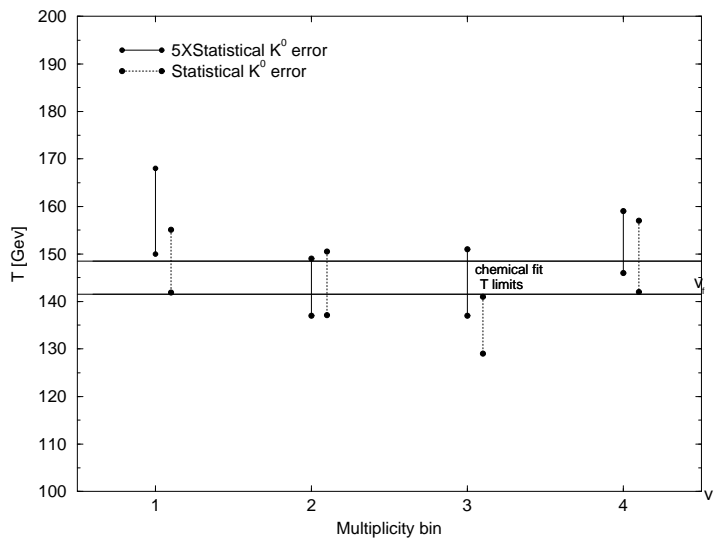}
}
\vspace*{-0.6cm}
\caption{ 
Thermal freeze-out temperature $T$ 
for different centrality bins compared to 
chemical freeze-out analysis shown by
horizontal solid lines. Original statistical error is used in the 
dotted results, five times statistical error for  kaon data is used 
in solid vertical lines, see text. 
\label{TdT}
}
\end{figure}
It is reassuring that we find a result
consistent with the purely chemical analysis of data
that included non-strange hadrons \cite{Let00}. 
There is 
no indication of a significant or systematic change of $T$ with centrality.
This is consistent with the believe that the formation of the new state of 
matter at CERN is occurring in all centrality bins explored by the 
experiment WA97. Only most peripheral interactions  
produce a change in the pattern of strange hadron production \cite{Kab99}. 
The (unweighted) average of all results shown in Fig.\,\ref{TdT}
produces a freeze-out temperature at the upper boundary of the 
the pure chemical freeze-out analysis result, $T\simeq 145$\,MeV. 
It should be noted that in chemical analysis 
$v_f=v$ \cite{Let00}, which may be the cause of this 
slight difference between current analysis average and the earlier 
purely chemical analysis result. 

The magnitudes of the 
collective expansion velocity $v$  and the  break-up (hadronization) speed 
parameter $v_f$ are presented in Fig.\,\ref{Tdv1v2}.
For $v$ (lower part of the figure) 
we again see consistency with earlier chemical freeze-out
analysis results, and there is no confirmed systematic trend
in the behavior of this parameter as function of centrality. 

Though within the experimental error, one could argue 
inspecting  Fig.\,\ref{Tdv1v2} that there is 
systematic increase in transverse flow velocity $v$ with centrality and thus 
size of the system. This is expected, since the more central events comprise 
greater volume of matter, which allows more time for development
of the flow.  Interestingly, it is in $v$ and not $T$ that we find the 
slight change of spectral slopes noted in the presentation of the 
experimental data \cite{Ant00}. 

The value of the break-up (hadronization) speed 
parameter $v_f=1/(\partial t_f/\partial r_f)$ 
shown in the top portion of Fig.\,\ref{Tdv1v2} is near to 
velocity of light which is highly  consistent with the picture of a 
sudden breakup of the  fireball. This 
hadronization surface velocity $v_f$ was in the earlier chemical
fit set to be equal to $v$, as there was not enough sensitivity in
purely chemical fit to  determine the value of $v_f$. 
\begin{figure}[tb]
\vspace*{-1.3cm}
\centerline{\hskip -0.9cm
\epsfig{width=10.cm,clip=,figure=\pathnow 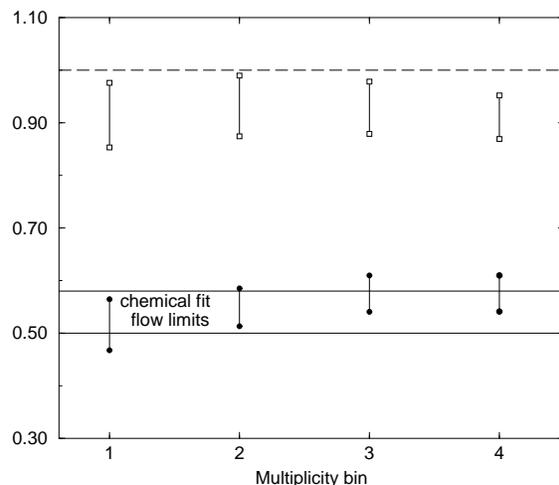}
}
\vspace*{-0.9cm}
\caption{ 
Thermal freeze-out flow velocity $v$ (top) and 
break up (hadronization) velocity $v_f$ 
for different centrality bins. Upper limit $v_f=1$ (dashed line) and 
chemical freeze-out analysis limits for $v$ (solid lines) are also shown.
\label{Tdv1v2}
}
\end{figure}

Unlike the temperature and two velocities, the overall normalization
of hadron yields, $V^h$ must be, and is strongly centrality dependent, as is
seen in Fig.\,\ref{Tdiagnorm}. This confirms in quantitative way the
believe that the entire available fireball volume is available for 
hadron production. The strong increase in the volume
by factor 6 is qualitatively 
consistent with a geometric interpretation of the 
collision centrality effect. Not shown is the error propagating from the
experimental data which is strongly correlated to the chemical
parameters discussed next. This systematic uncertainty is another reason 
we do not attempt an absolute unit volume normalization. 

The 4 chemical parameters $\lambda_q,\lambda_s, \gamma_q, \gamma_s/\gamma_q$ are
shown in the following Figures \ref{Tdlqls},\ref{Tdgamqgams}.
These parameters determine along with $V^h$ the final particle
yield. Since we have 5 parameters 
determining normalization of  6 strange hadron spectra, and as discussed we 
reduce the statistical weight of Kaons, there
is obviously a lot of correlation between these 4 quantities, and thus 
the error bar which reflects this correlation,  is significant. 

The chemical fugacities $\lambda_q$ and $\lambda_s$ shown in 
Fig.\,\ref{Tdlqls} do not exhibit
a systematic centrality dependence. This is consistent with the result
we found for $T$ in that the freeze-out properties of the fireball are
seen to be for the temperature and chemical potential values independent
of the size of the fireball. Comparing to the earlier chemical
freeze-out result in Fig.\,\ref{Tdlqls} one may argue that 
there is a systematic downward deviation in $\lambda_q$. However,
 this could be  caused by the fact that the 
chemical freeze-out analysis allowed for
isospin-asymmetric $\Xi^-(dss)$ yield \cite{Let00}, 
while out present analysis is not
yet distinguishing light quarks.

\begin{figure}[tb]
\vspace*{-1.3cm}
\centerline{\hskip 0.5cm
\epsfig{width=9.5cm,clip=,angle=-90,figure=\pathnow 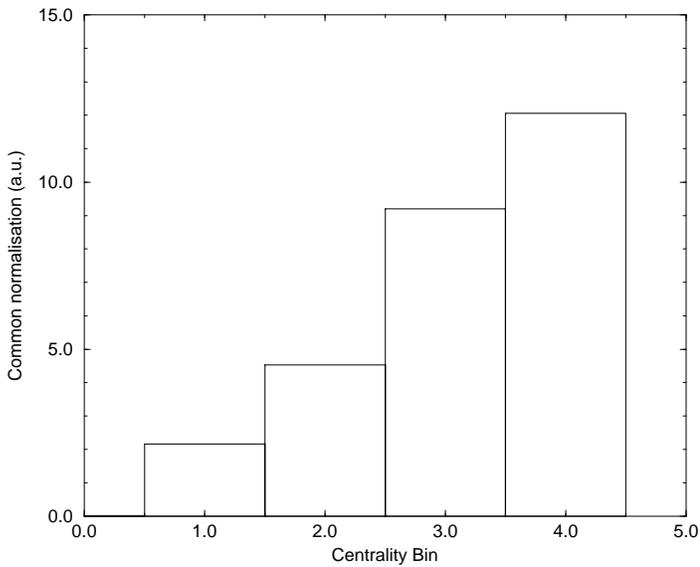}
}
\vspace*{-0.2cm}
\caption{ 
Hadronization volume (arbitrary units)
 for different centrality bins. 
\label{Tdiagnorm}
}
\end{figure}

\begin{figure}[tb]
\vspace*{-2.9cm}
\centerline{\hskip -2.5cm
\epsfig{width=11.cm,clip=,angle=-90,figure=\pathnow 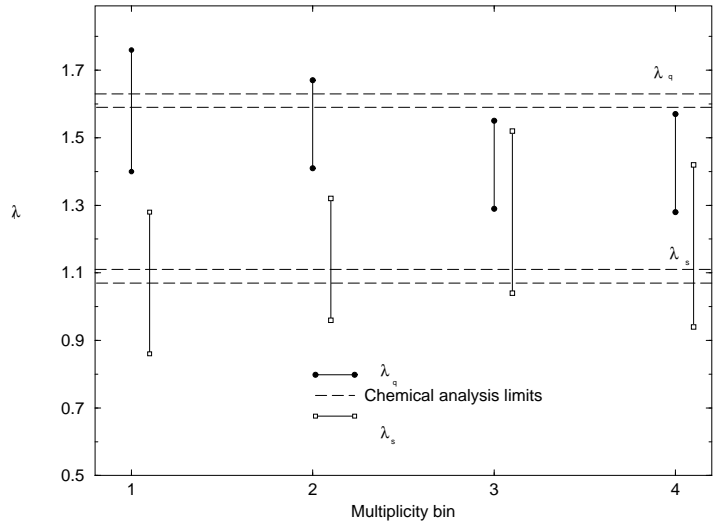}
}
\vspace*{-0.2cm}
\caption{ 
Thermal analysis chemical quark fugacity $\lambda_q$ (top) and 
strange quark fugacity $\lambda_s$ (bottom)
 for different centrality bins compared to 
the chemical freeze-out analysis results. 
\label{Tdlqls}
}
\end{figure}

The ratio $\gamma_s/\gamma_q$  shown in bottom portion of  Fig.\,\ref{Tdgamqgams}
 is systematically smaller than unity, consistent with many years of prior analysis:
when $\gamma_q=1$ is tacitly chosen, this ratio is the value of $\gamma_s$
in analysis of strange baryons.
We have not imposed a constrain on the range of $\gamma_q$ 
(top of Fig.\,\ref{Tdgamqgams}) and thus 
values greater than the pion condensation point 
$\gamma_q^*=e^{m_\pi/2T}\simeq 1.65$ (thick line) can be expected, but in fact 
do not arise.

It is important to explicitly check
how well the particle $m_\bot$-spectra are
reproduced. We group all bins in one figure and show 
in Figs.\,\ref{TdLamtot},
\ref{TdALamtot}, \ref{TdXitot}, \ref{TdAXitot} 
in sequence  $\Lambda,\,\overline\Lambda,\,\Xi,\,\overline\Xi$. 
It is important to note that there are some significant 
deviations which appear to be falling outside of the trend set by the 
other measurements. --this  occurs for $\Lambda$ as well 
but remains invisible in the figure due to the smallness of the 
experimental error bar. Overall, the description of the shape of the spectra 
is very satisfactory.

\begin{figure}[tb]
\vspace*{-2.5cm}
\centerline{\hskip -1.cm
\epsfig{width=9.5cm,height=9.5cm,clip=,angle=-90,figure=\pathnow 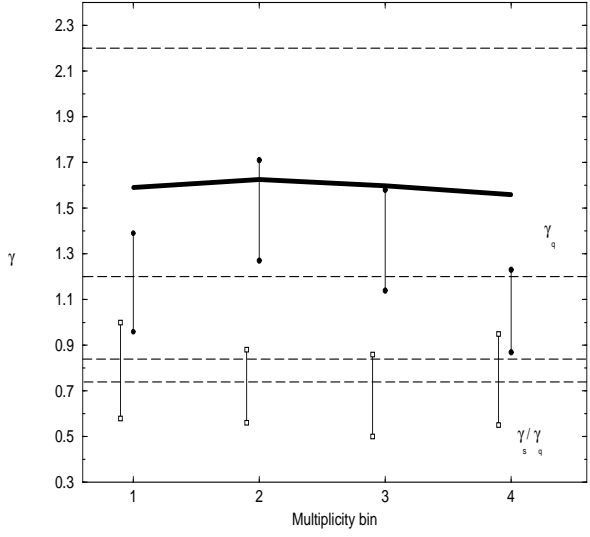}
}
\vspace*{.7cm}
\caption{ 
Thermal analysis chemical quark abundance parameter $\gamma_q$ (top)
and $\gamma_s/\gamma_q$ (bottom)  for different centrality bins compared to 
the chemical freeze-out analysis. Thick line: upper limit due to pion
condensation.
\label{Tdgamqgams}
}
\end{figure}

\begin{figure}[tb]
\vspace*{-1.8cm}
\centerline{\hskip 0.5cm
\epsfig{width=10.cm,clip=,angle=-90,figure=\pathnow 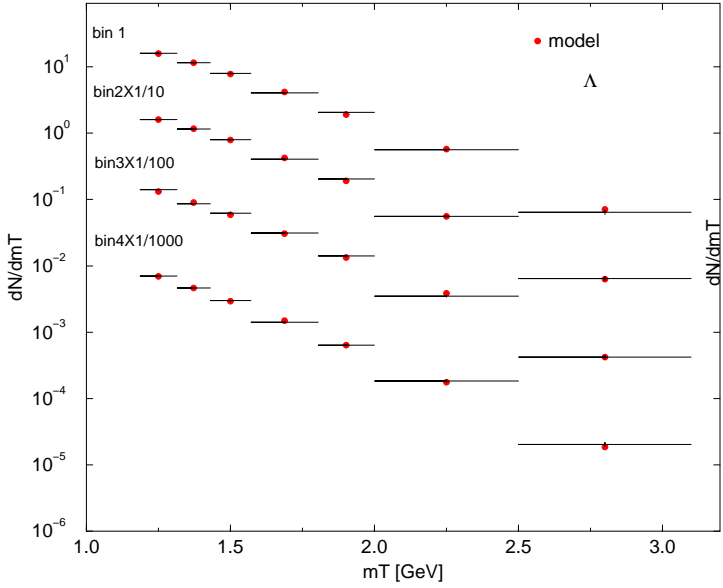}
}
\vspace*{-0.2cm}
\caption{ 
Thermal analysis $m_T$ spectra: $\Lambda$ 
\label{TdLamtot}
}
\end{figure}

We also describe the K$_0$data extremely well,
especially 
in the $m_\bot$ range which is the same as that for
hyperons considered earlier,  as is seen in Fig.\,\ref{TdK0All}.
We recall that these results were obtained reducing the 
statistical significance of Kaon data, and thus the conclusion
is that hyperons predict both the abundance and shape of 
kaon spectra. Moreover, all the strange hadron spectra can be
well described within the model we have adopted.  

\begin{figure}[tb]
\vspace*{-1.5cm}
\centerline{\hskip 0.5cm
\epsfig{width=10.cm,clip=,angle=-90,figure=\pathnow 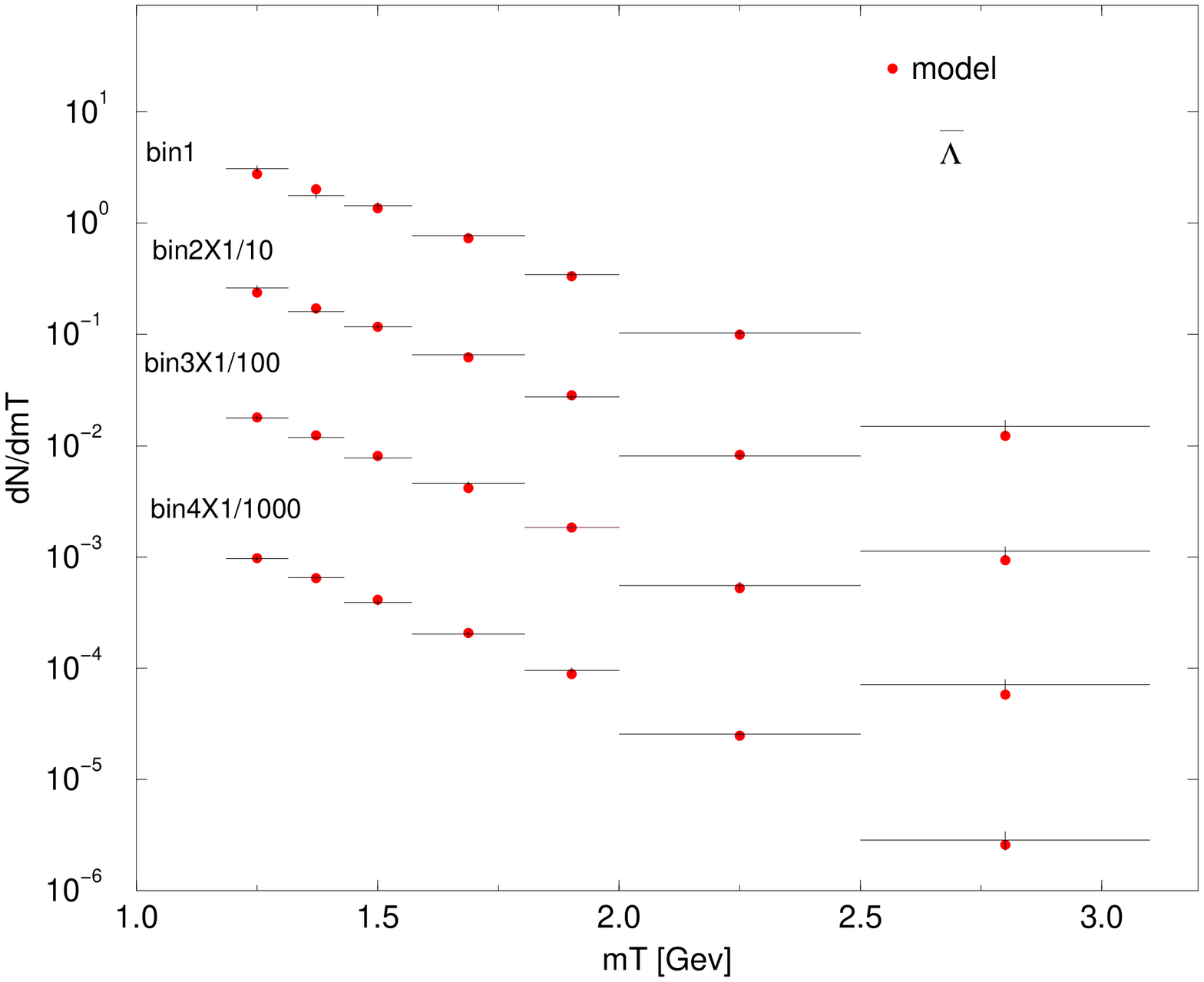}
}
\vspace*{-0.2cm}
\caption{ 
Thermal analysis $m_T$ spectra: $\overline\Lambda$. 
\label{TdALamtot}
}
\end{figure}

\begin{figure}[tb]
\vspace*{-1.5cm}
\centerline{\hskip 0.5cm
\epsfig{width=10.cm,clip=,angle=-90,figure=\pathnow 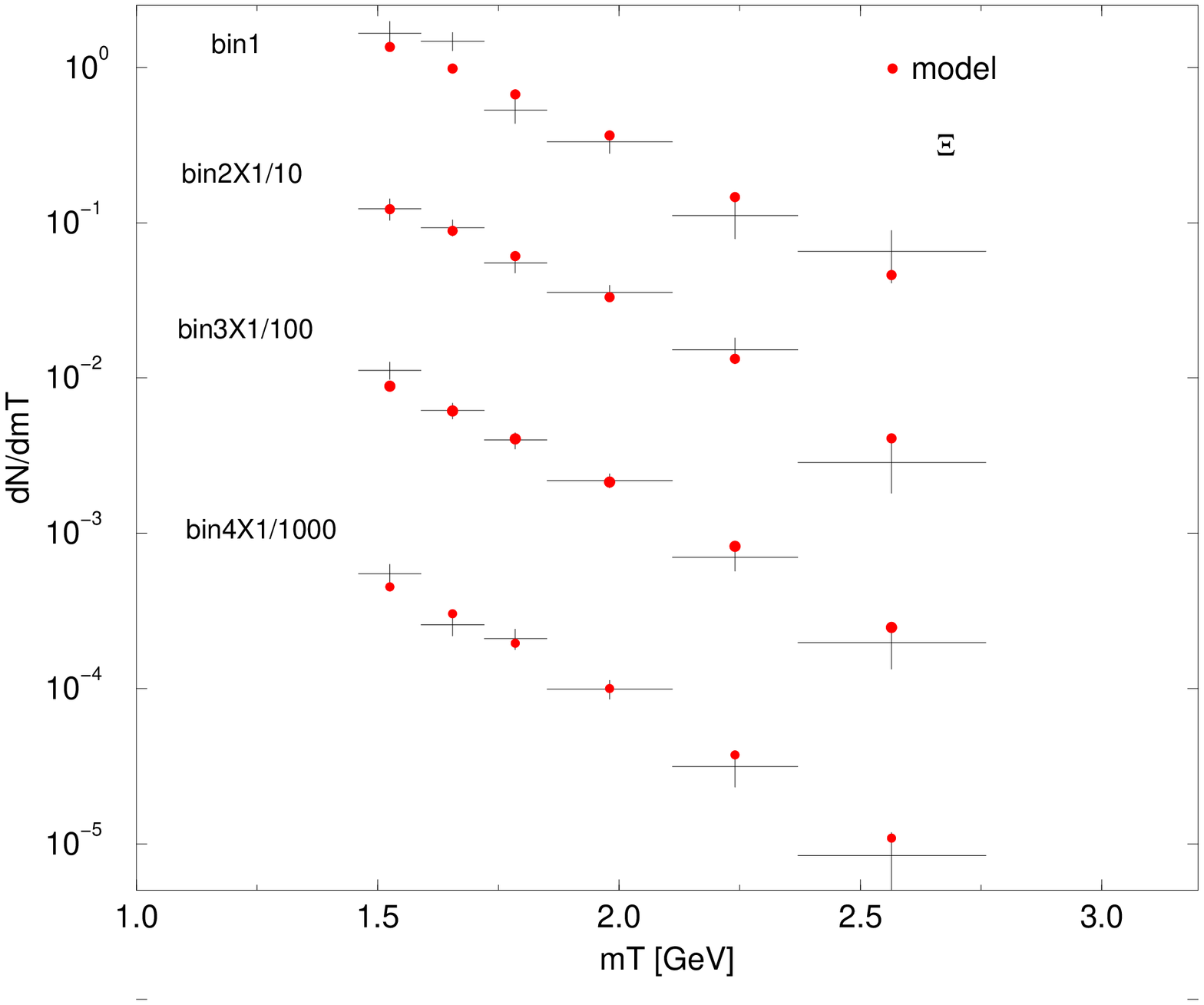}
}
\vspace*{-0.2cm}
\caption{ 
Thermal analysis $m_T$ spectra: $\Xi$
\label{TdXitot}
}
\end{figure}

\begin{figure}[tb]
\vspace*{-1.5cm}
\centerline{\hspace*{0.70cm}
\epsfig{width=10cm,clip=,angle=-90,figure=\pathnow 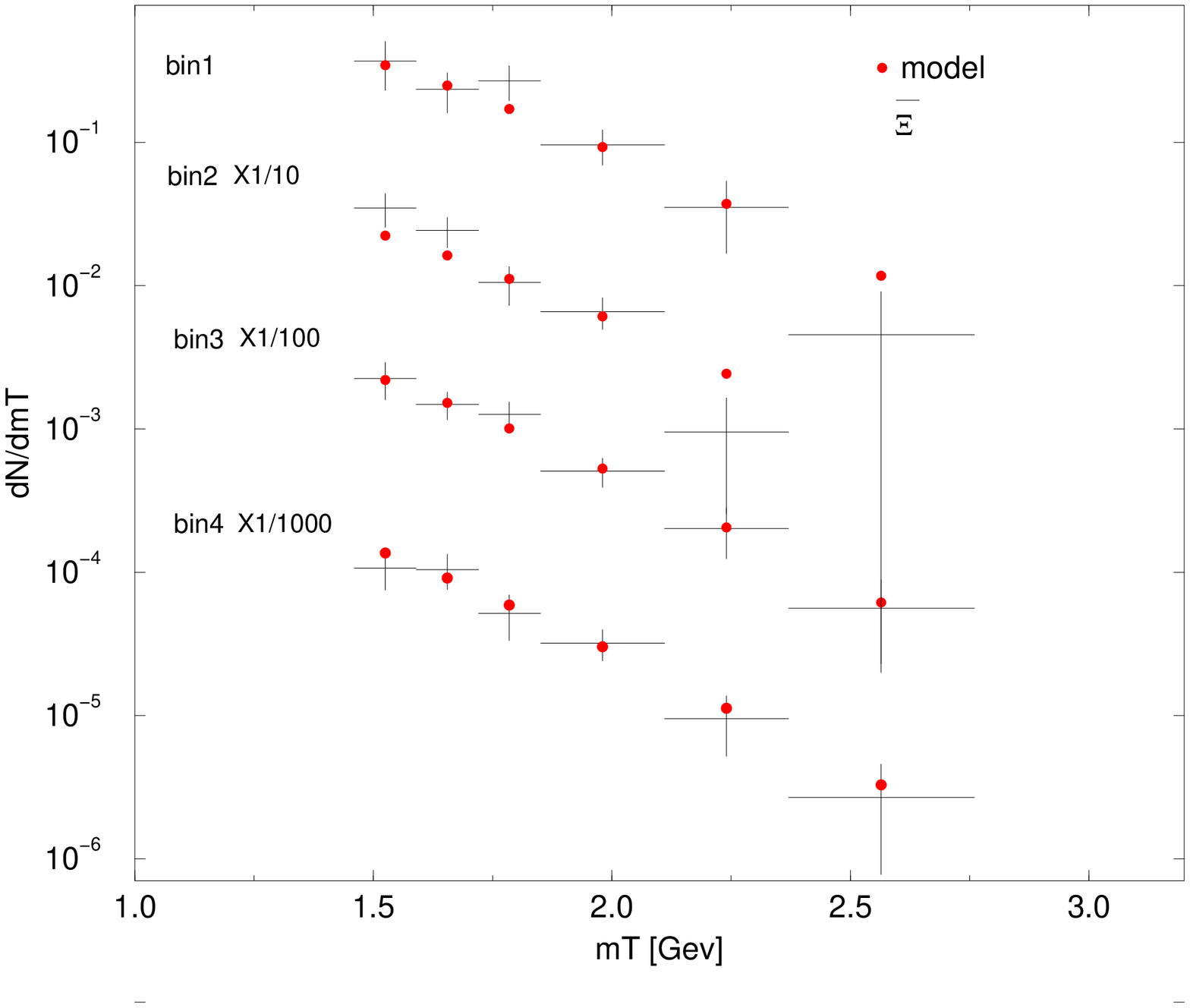}
}
\vspace*{-0.2cm}
\caption{ 
Thermal analysis $m_T$ spectra: $\overline\Xi$
\label{TdAXitot}
}
\end{figure}

\begin{figure}[tb]
\vspace*{-1.2cm}
\centerline{\hskip 0.5cm
\epsfig{width=10.cm,clip=,angle=-90,figure=\pathnow 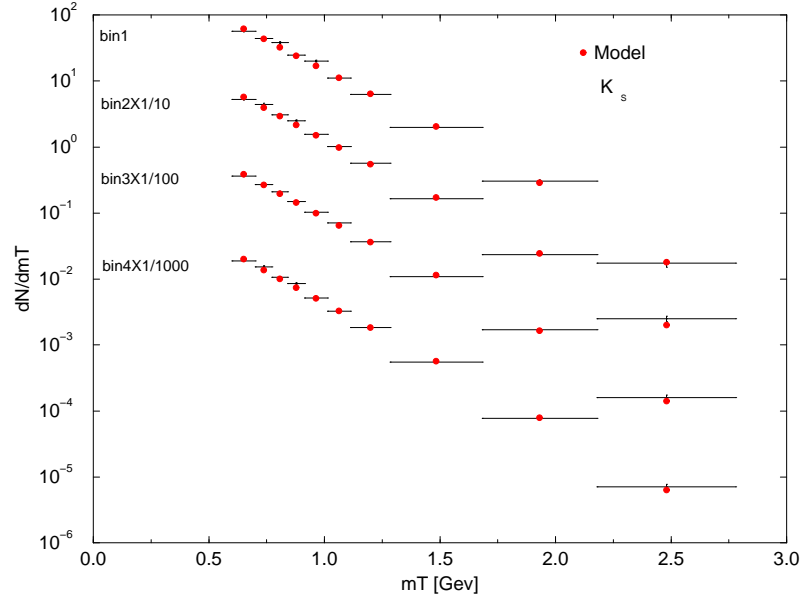}
}
\vspace*{-0.2cm}
\caption{ 
Thermal analysis $m_T$ spectra: $K_s$.
\label{TdK0All}
}
\end{figure}

\section{Statistical significance of the 
results presented}\label{chidata}
We have analysed the validity and consistency 
of our data analysis by exploring $\chi^2/DoF$ 
profiles. These are obtained by fixing the 
value of one of the parameters (we consider $T_f,
v, \partial t_f/\partial r_f=1/v_f$), and computing 
the related  $\chi^2_{\rm T}/DoF$, the total error 
divided by degrees of freedom. These are the number 
of measurements minus number of parameters, $DoF$ is
typically 33 in this data analysis.  All curves must have 
the same $\chi^2_{\rm T}/DoF$ at the minimum and this 
minimum must point to the value of parameters we report. 
For the temperature $T$ we produced two results 
shown in figure \ref{TchiT2}, in the bottom section
for experimental (statistical) $K^0$ measurement error,
and in the top part for the five times enlarged error.
We recall that both results are presented in Fig.\,\ref{TdT}.
We note that there is a pronounced $\chi^2_{\rm T}/DoF$ 
minimum shown on logarithmic scale) 
for all 8  results of which the average value is at 
$T=145$ MeV. 

We show in Fig.\,\ref{Tchivv2} the 
profile of $\chi^2_{\rm T}/DoF$  for the collective flow velocity
v (top) and the freeze-out surface 
$\partial t_f/\partial r_f=1/v_f$ motion (bottom) being 
fixed. These minima can be shown on linear scale. We note
a mild secondary minimum in the region 
$v\simeq 0.25$--0.35. However, the
minima we find at $v=0.5$--0.58 are by far more significant.
$\partial t_f/\partial r_f=1/v_f$ is converging to a sharp
minimum seen in bottom portion of  Fig.\,\ref{Tchivv2}
at at a value consistent with the sudden breakup scenario.
It is necessary to include $\partial t_f/\partial r_f=1/v_f$
along with $v$ in the analysis to find this result, 
which was not always done in other studies of particle spectra.

\begin{figure}[tb]
\vspace*{-1.5cm}
\centerline{
\epsfig{width=9.5cm,clip=,angle=-90,figure=\pathnow 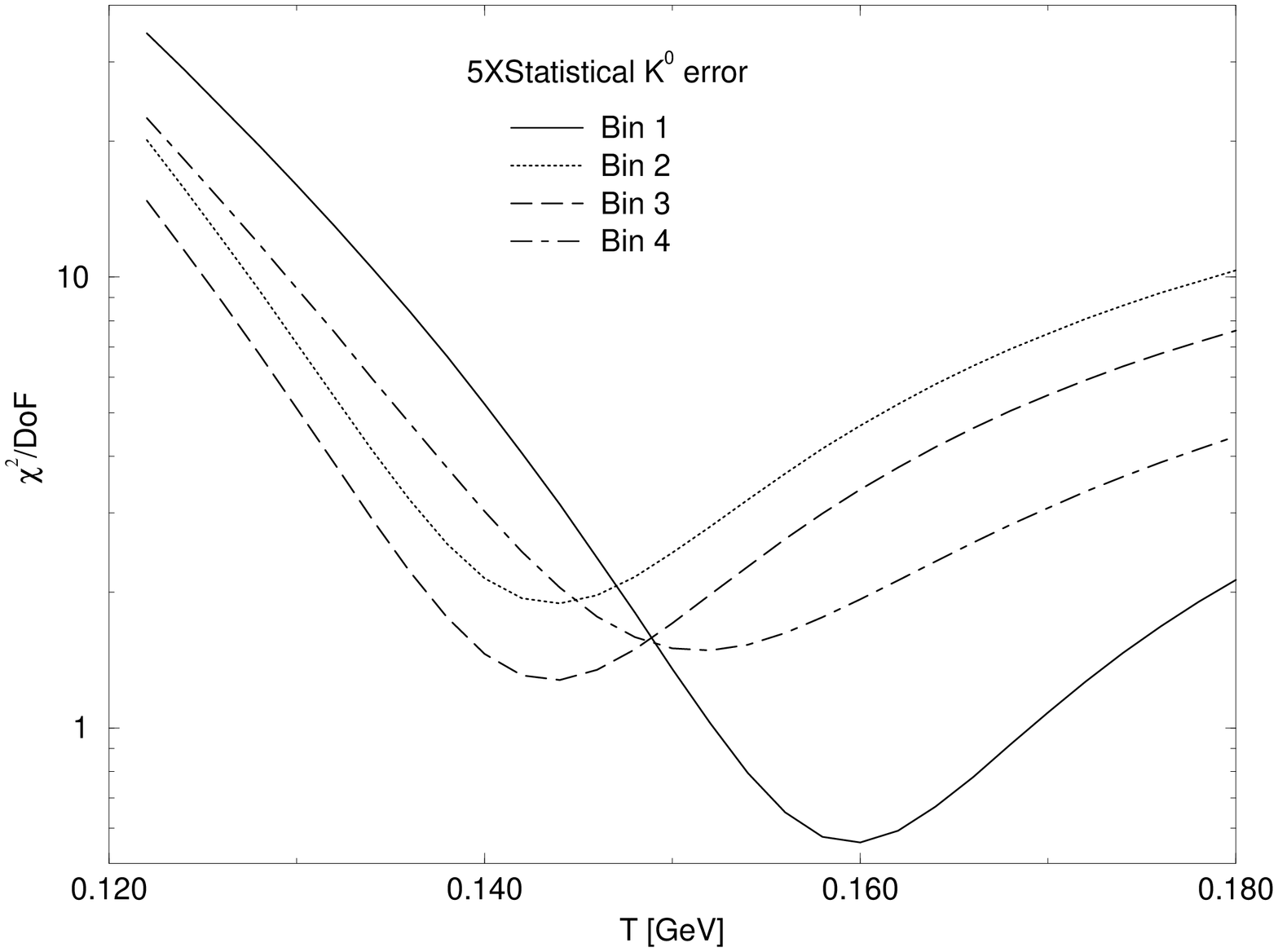}
}
\vspace*{-1.7cm}
\centerline{
\epsfig{width=9.5cm,clip=,angle=-90,figure=\pathnow 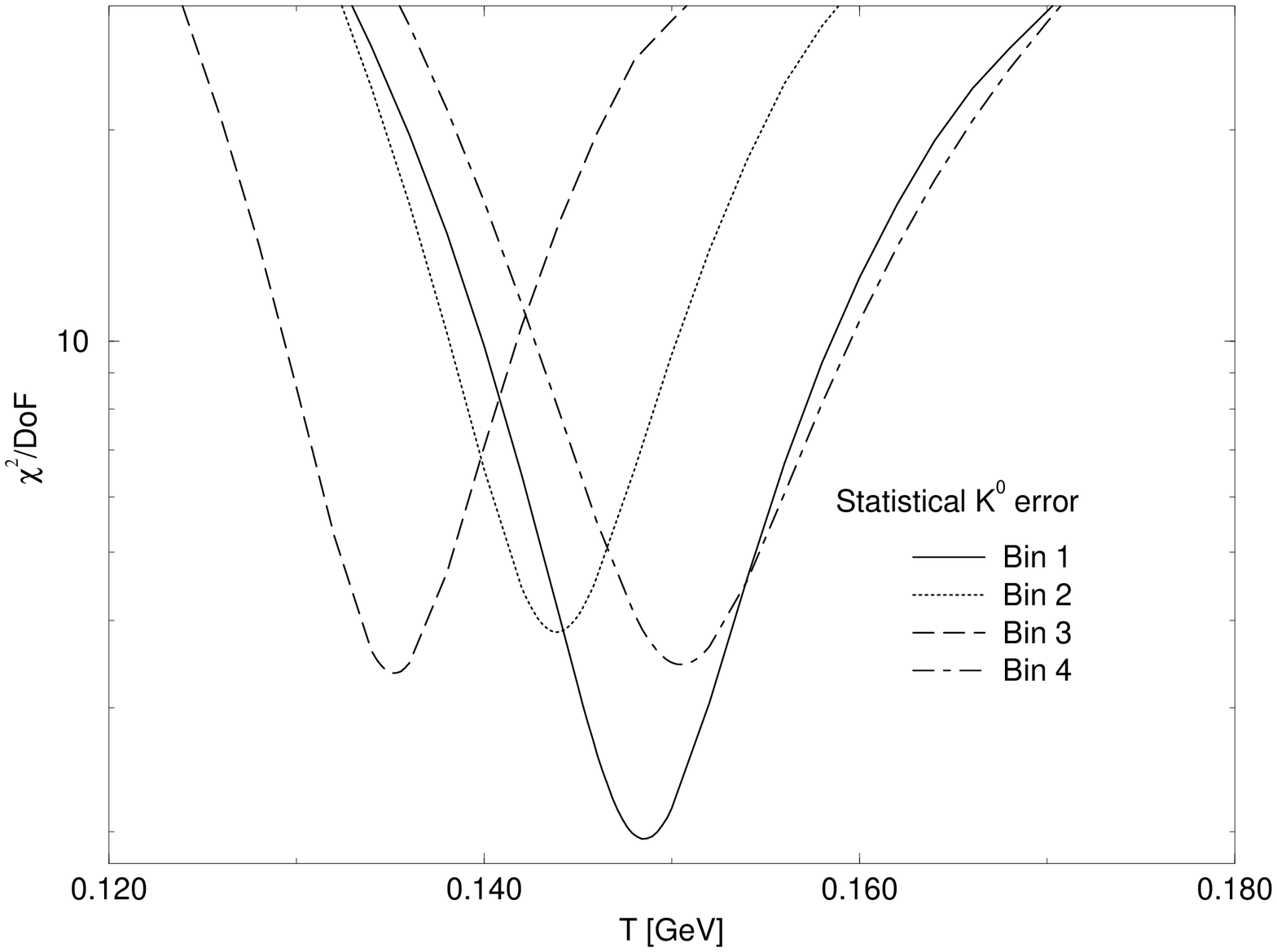}
}
\vspace*{-.3cm}
\caption{ 
The total error divided by degrees of freedom for different
 centrality bins, shown as function of (fixed) 
freeze-out temperature $T$, bottom for the  experimental
value of the (statistical) $K^0$ error, top for 
the 5 times enlarged kaon data statistical errors.
\label{TchiT2}
}
\end{figure}

\section{Omega spectra}
In Fig.\, \ref{TdOAOtot}
all four centrality bins for the sum $\Omega+\overline\Omega$ are
shown. We see that we systematically  under predict the two lowest
$m_\bot$ data points. Some deviation at high $m_\bot$ may be attributable
to acceptance uncertainties, also seen in the the 
$\Xi$ result presented earlier in Fig.\,\ref{TdAXitot}. 
 We recall that there is a disagreement with the 
Omega yields in the chemical analysis, which thus does
not include in the analysis the production of $\Omega$. In the here presented 
analysis we see that this disagreement is arising at
low momentum.

The low-$m_\bot$ anomaly also explains why the 
inverse $m_\bot$ slopes for $\Omega,\overline\Omega$ 
are smaller than the values seen in all other strange (anti)hyperons.
One can presently only speculate about the processes 
which contribute to this anomaly. 
We note that the 1--2s.d. deviations in the 
low  $m_\bot$-bins of the $\Omega+\overline\Omega$ 
spectrum translates into 3s.d. 
deviations from the prediction of the chemical analysis. 
This is mainly a consequence of the fact that after summing
over centrality and $m_\bot$, the statistical error which dominates
$\Omega+\overline\Omega$ spectra becomes relatively small, and
as can be seen the low $m_\bot$ excess practically doubles
the $\Omega$ yield. 

\begin{figure}[tb]
\vspace*{-1.3cm}
\centerline{
\epsfig{width=9.5cm,clip=,angle=-90,figure=\pathnow 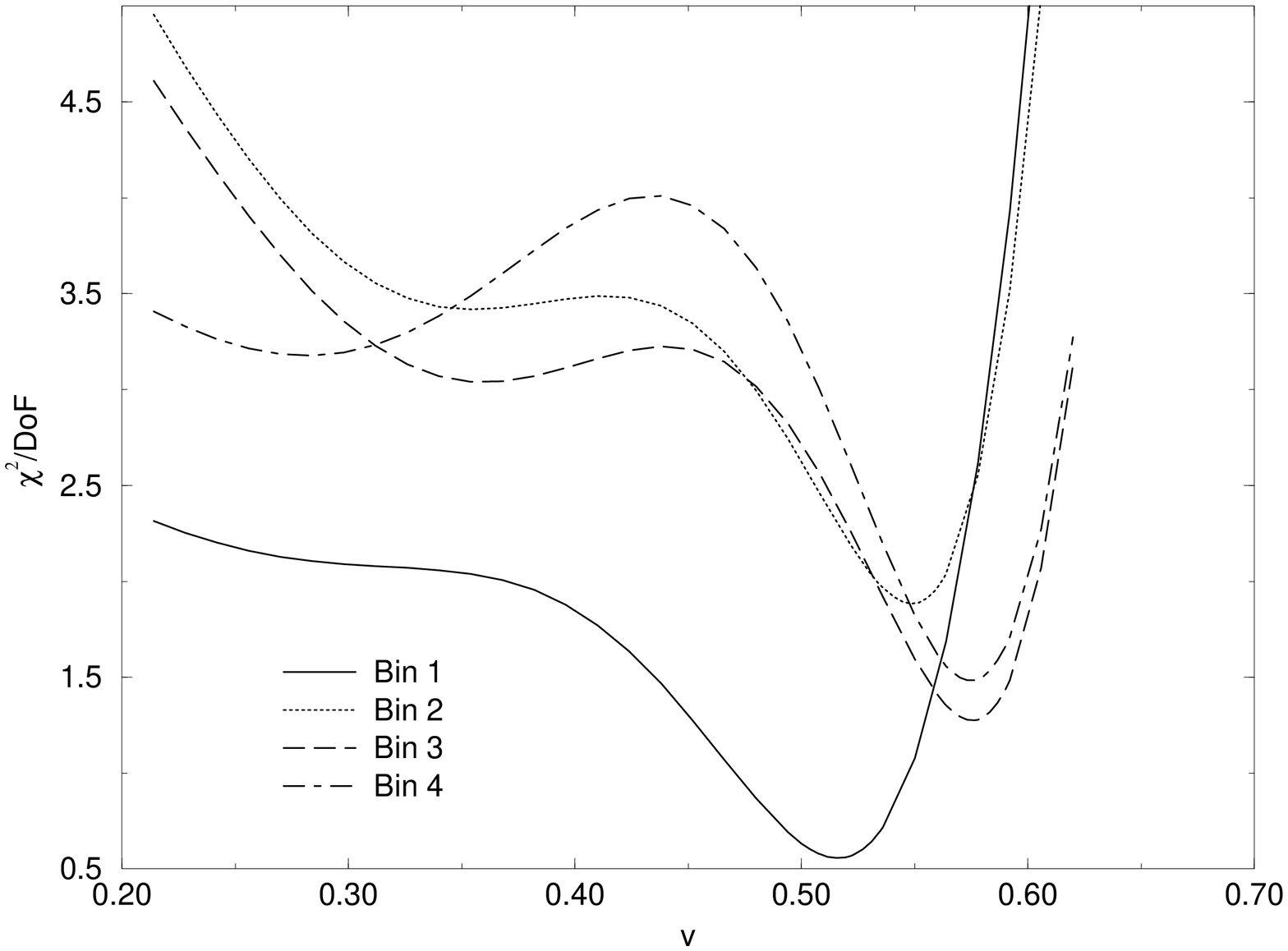}
}
\vspace*{-1.7cm}
\centerline{
\epsfig{width=9.5cm,clip=,angle=-90,figure=\pathnow 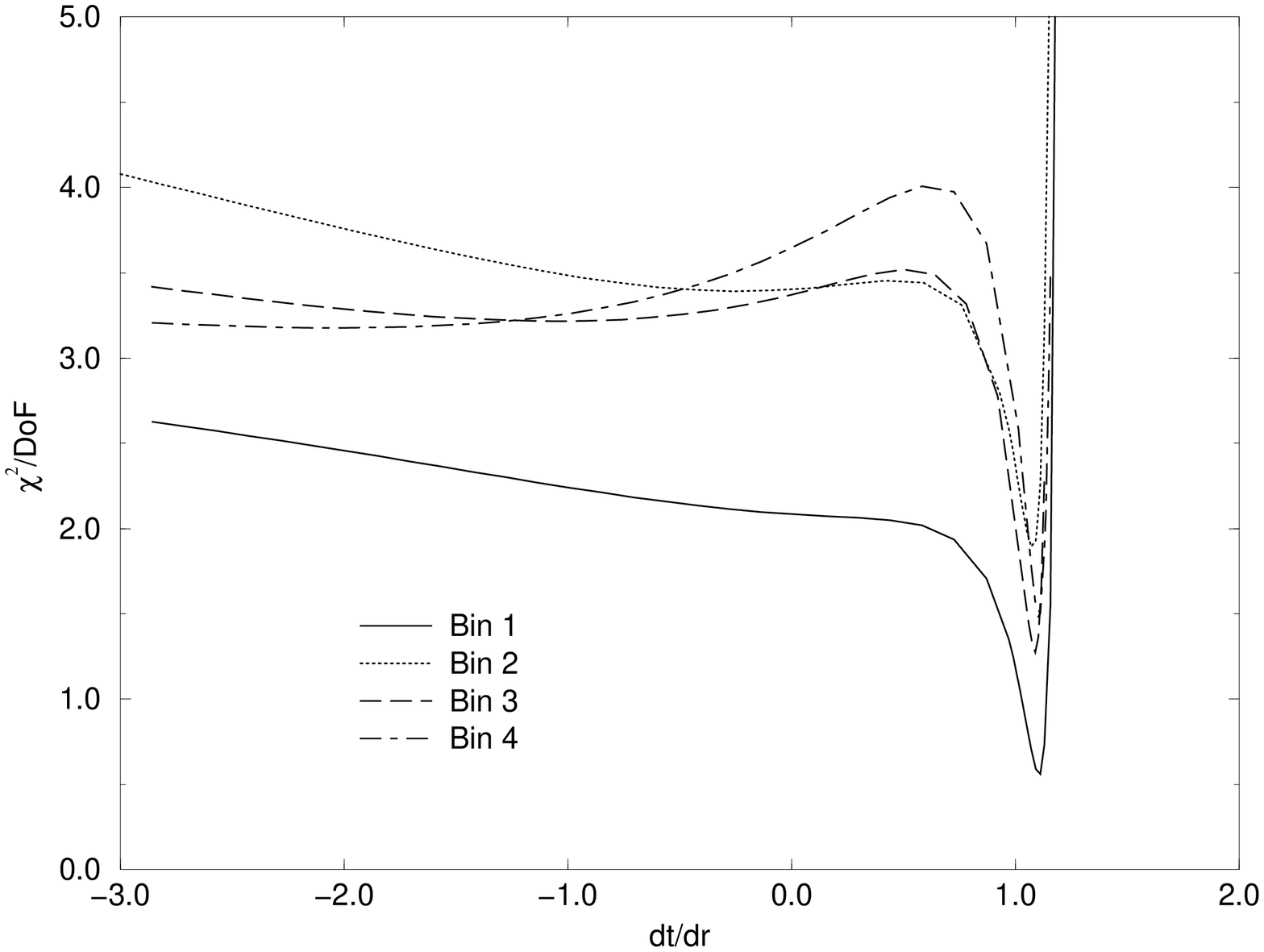}
}
\vspace*{-.3cm}
\caption{ 
The total error divided by degrees of freedom  for different
 centrality bins, shown as function of (fixed) flow velocity  
 $v$ on top and for (fixed) freeze-out surface 
$\partial t_f/\partial r_f=1/v_f$ dynamics on the bottom. 
\label{Tchivv2}
}
\end{figure}

\begin{figure}[tb]
\vspace*{-1.5cm}
\centerline{
\epsfig{width=9.5cm,clip=,angle=-90,figure=\pathnow 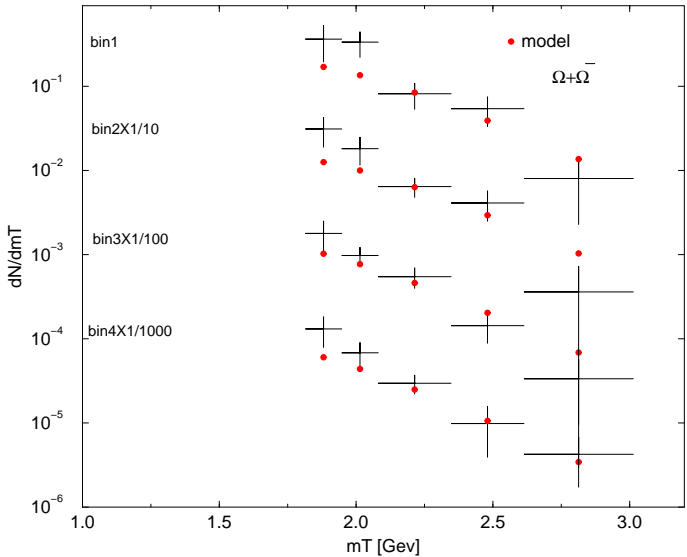}
}
\vspace*{-0.6cm}
\caption{ 
Thermal analysis $m_T$ spectra: $\Omega+\overline\Omega$.
\label{TdOAOtot}
}
\end{figure}

\section{Final remarks}
Our thermal freeze-out analysis confirms that
CERN-SPS  results decisively show  interesting and new 
physics, and confirms the reaction picture of
a suddenly hadronizing QGP-fireball with both
chemical and thermal freeze-out being the same. 
In our view driven
by internal pressure, a  quark-gluon fireball expands and 
ultimately  a sudden breakup (hadronization) 
into final state  particles occurs which reach detectors 
without much, if any, further rescattering. The required
sudden fireball breakup  arises as the fireball super-cools, and in this 
state encounters a strong mechanical instability \cite{Raf00}. Note that
deep super cooling requires a  first order phase transition.

The remarkable similarity of 
$m_\bot$ spectra reported by  the  WA97 experiment
is interpreted by a set of freeze out parameters,
and we see that production mechanism of 
$\Lambda$, $\overline{\Lambda}$, and  $\Xi$, 
$\overline\Xi$ is the same. This symmetry, including matter--antimatter 
production is an important cornerstone of the claim that the strange
antibaryon data can only be interpreted in terms of direct particle emission
from a deconfined phase.

The reader must remember that  in presence of conventional 
hadron collision based physics, the production mechanism 
of antibaryons is quite different from that of baryons 
and a similarity of the $m_\bot$ spectra is not 
expected. Moreover, even if QGP is formed, but a equal
phase of confined particles is present, the annihilation of 
antibaryons in the baryon rich medium created at CERN-SPS energy
would deplete more strongly antibaryon yields, in
particular so  at small particle momentum, 
with the more abundant baryons remaining less influenced. 
This effect is not observed \cite{Ant00}. 

Similarity of $m_\bot$-spectra 
does not at all imply in our argument a similarity of particle rapidity spectra.
As hyperon are formed at the fireball breakup, any remaining longitudinal
flow present among fireball constituents 
will be imposed on the product particle, thus $\Lambda$-spectra
 containing potentially two original valence quarks are stretched in
$y$, which  $\overline\Lambda$-$y$-spectra are not, as they are made from newly 
formed particles. All told, one would expect that anti-hyperons can
 appear with a thermal
rapidity distribution, but hyperons will not. But both have the same 
thermal-explosive collective flow controlled shape of $m_\bot$-spectra.

We have shown  that  thermal freeze-out 
condition for strange hadrons (K$^0_s, \Lambda, \overline\Lambda,
\Xi, \overline\Xi$)  agrees within error with chemical freeze-out
and we have confirmed the  freeze-out temperature $T\simeq 145$\,MeV.
These findings about the similarity of thermal and chemical
freeze-out were controversial, when   the experimental single 
particle spectra  were lacking precision, since pion spectra and
two particle correlation analysis did not yield this result. 
However, this paper
studies the  precise hyperon and kaon  $m_\bot$ spectra
which reach to relatively low $p_\bot$ and compares 
with definitive chemical analysis of SPS data. The two particle
correlation analysis involves pions, which unlike strange
hadrons here considered, are potentially witnesses to other 
physics than the properties of dense and hot quark-gluon phase.

We were able to determine the freeze-out surface
$1/v_f=\partial t_f/\partial r_f$
dynamics and have shown that the break-up velocity $v_f$ is nearly
the velocity of light, as would be expected in  a sudden breakup of a
QGP fireball.  A study with $\partial t_f/\partial r_f$  has not
been previously considered, and only collective flow is included in
the description of the particle source. In our analysis we 
find a slight increase of the transverse expansion 
velocity with the size of the fireball volume, 
but consistently $v\le 1/\sqrt{3}$. 

We have reproduced the strange particle spectra in all centrality bins. 
Our findings rely strongly on
results obtained by WA97 at smallest accessible  particle momentum, 
and this stresses the need to reach to smallest possible
$p_\bot$ in order to be able to explore the physics of particle
freeze-out from the deconfined region.  Moreover, we 
demonstrated that the experimental production data 
of $\Omega+\overline\Omega$ 
has a noticeable systematic low $p_\bot$ enhancement anomaly 
present in all centrality bins. This result shows that it is not a
different temperature of freeze-out  of $\Omega+\overline\Omega$ 
 that leads to more enhanced yield, but a soft momentum secondary source
which contributes almost equal number of soft $\Omega+\overline\Omega$
compared to the systematic yield predicted by the other strange hadrons. 

{\vspace{0.5cm}\noindent\it Note added:\\}
We have been made aware by the referee that an
analysis of the $m_\bot$ spectra for high energy collisions 
has been carried out recently \cite{Bec00}. This work 
reaches for elementary high energy processes similar conclusions 
as we have presented regarding the identity of 
chemical and thermal freeze-out. The higher 
freeze-out temperature 
found is also consistent with our results, considering that 
in nuclear collisions significant 
super cooling occurs \cite{Raf00}. 

{\vspace{0.5cm}\noindent\it Acknowledgments:\\}
Supported  by a grant from the U.S. Department of
Energy,  DE-FG03-95ER40937\,. 

\vskip -0.7cm

\end{narrowtext}
\end{document}